\documentclass[12pt]{article}
\usepackage{amssymb}
\usepackage{epsfig}
\usepackage{graphicx}
\usepackage{amsmath}
\usepackage{verbatim}

\csname @addtoreset\endcsname{equation}{section}
\begin{document}
\begin{titlepage}
\title{\bf\Large Scalar Explanation of Diphoton Excess at LHC \vspace{18pt}}

\author{\normalsize Huayong Han, Shaoming Wang and Sibo Zheng \vspace{12pt}\\
{\it\small   Department of Physics, Chongqing University, Chongqing 401331, P.R. China}\\
}

\date{}
\maketitle \voffset -.3in \vskip 1.cm \centerline{\bf Abstract}
\vskip .3cm
Inspired by the diphoton signal excess observed in the latest data of 13 TeV LHC,
we consider either a 750 GeV real scalar or pseudo-scalar responsible for this anomaly.
We propose a concrete vector-like quark model,
in which the vector-like fermion pairs directly couple to this scalar via Yukawa interaction.
For this setting the scalar is mainly produced via gluon fusion, 
then decays at the one-loop level to SM diboson channels 
$gg, \gamma\gamma, ZZ, WW$.
We show that for the vector-like fermion pairs with exotic electric charges, 
such model can account for the diphoton excess
and is consistent with the data of 8 TeV LHC simultaneously in the context of perturbative analysis.

\end{titlepage}
\newpage

\section{Introduction}\label{intro}
The first data at the 13 TeV Large Hadron Collider (LHC)
was released on December 15 2015 \cite{13tevatlas,13tevcms}.
It shows an excess in diphoton final state 
at the invariant mass $M\simeq750$ GeV,
with local significance of order 3.9 $\sigma$ and 2.6 $\sigma$ for ATLAS and CMS, respectively.
In contrast, no excesses in the Standard Model (SM) diboson channels such as $\gamma\gamma, ZZ,WW, ZW$, dilepton and dijet 
were seen in the old data of 8 TeV LHC \cite{1506.02301,8tevcms, 8tevatlas,
8tev-diphoton, 8tev-zz,8tev-ww, 8tev-dijet, 8tev-bb, 8tev-tt}.

If the diphoton excess is indeed a hint of some new physics beyond SM, 
for an on-shell decay to diphoton it should be due to either spin-0 or spin-2 scalar $\phi$.
To explain the observed excess,
the cross section $\sigma(pp\rightarrow \phi\rightarrow \gamma\gamma)$ is required to satisfy the signal strength of order,
\begin{eqnarray}\label{excess}
\sigma(pp\rightarrow \phi\rightarrow \gamma\gamma)\mid_{\sqrt{s}=13~\text{TeV}}
\simeq (8\pm 3)~\text{fb}.
\end{eqnarray}
Such SM singlet scalar which is responsible for the excess 
has stimulated extensive interests, 
see Ref.\cite{1512.04850}- Ref.\cite{1512.06113}.
  
In this paper, we propose a concrete vector-like quark model,
in which the vector-like fermion pairs directly couple to $\phi$ 
via tree-level Yukawa interaction. 
Under our setup, $\phi$ is mainly produced via gluon fusion, 
then decays at the one-loop level to SM diboson channels $gg,γγ,ZZ,WW$, 
with the colored vector-like fermion pair running in the Feynman loop. 
For the vector-like fermion pairs with exotic electric charges, 
such model can account for the diphoton excess,
and is consistent with the data of 8 TeV LHC simultaneously in the context of perturbative analysis.

This paper is organized as follows.
In Sec.2 we address the matter content in the vector-like quark model,
define the parameter space,
and summarize the experimental limits on $\phi$ and vector-like quark at the 8 TeV LHC.
In Sec.3 we explore the parameter space for $\phi$ 
either being a real scalar or pseudo-scalar.
Finally, we conclude in Sec. 4.

\section{The Vector-like Quark Model}
\subsection{The Model}
 \begin{table}
\begin{center}
\begin{tabular}{|c|c|c|c|c|}
  \hline
 \text{Matters} &  $SU(3)_{c}$& 
$SU(2)_{L}$ & $U(1)_{Y}$   \\
  \hline\hline
  $\phi$ & $\mathbf{1}$  &  $\mathbf{1}$ & $0$\\
\hline 
$\Psi=(\psi_{1},\psi_{2})^{T}$& $\mathbf{3}$  & $\mathbf{2}$  &  $q_{\psi}$ \\
\hline
$\tilde{\Psi}=(\tilde{\psi}_{1},\tilde{\psi}_{2})^{T}$&  $\bar{\mathbf{3}}$ & $\bar{\mathbf{2}}$  & -$q_{\psi}$  \\
\hline
 \end{tabular}
\caption{Matters and their SM quantum numbers in the vector-like quark model. Another fermion doublet $\tilde{\Psi}$ is added to make sure that the model is free of gauge anomaly. }
\label{table}
\end{center}
\end{table}

In order to reproduce the on-shell decay $\phi\rightarrow \gamma\gamma$,
which is a loop process for the SM singlet $\phi$,
we directly couple $\phi$  to a fermion doublet $\Psi$, 
the latter of which is a subsector of vector-like quark model as defined in Table \ref{table}. 
In this table, another fermion doublet $\tilde{\Psi}$ is added in order to evade the gauge anomaly problem.

For simplicity, we assume that the mass $M_{\tilde{\Psi}}$ for $\tilde{\Psi}$ is obviously larger than the mass $M_{\Psi}$ for $\Psi$.
Below the mass scale $M_{\tilde{\Psi}}$ the effective Lagrangian in the new physics is described by 
\footnote{The effective Lagrangian as analyzed in the previous version of this manuscript is a simplification of this concrete one.},
\begin{eqnarray}\label{Lagrangian}
\mathcal{L}_{\text{BSM}}=\frac{1}{2} \left(\partial \phi\right)^{2}-\frac{1}{2}m_{\phi}^{2}\phi^{2}
+ i\bar{\Psi} \gamma^{\nu}{D}_{\nu}\Psi -M_{\Psi}\bar{\Psi}\Psi+\mathcal{L}_{\text{Yukawa}} ,
\end{eqnarray}
where 
\begin{eqnarray}\label{interaction}
\mathcal{L}_{\text{Yukawa}}=
 \begin{cases}
y\phi\bar{\Psi}\Psi, &  (\text{scalar}), \\
iy\phi\bar{\Psi}\gamma_{5}\Psi, &  (\text{pseudo-scalar}).
\end{cases}
\end{eqnarray}
In Eq.(\ref{Lagrangian}) scalar mass $m_{\phi}\simeq 750$ GeV,
$y$ is the Yukawa coupling constant.
We assign the electric charge for $\Psi$ as $Q_{\psi_{1,2}}=q_{\psi}\pm \frac{1}{2}$ in unit of $e$.
For either case in Eq.(\ref{interaction}) 
$\phi$ is mainly produced by gluon fusion,
and decays to diphoton via $\Psi$ in the Feynman loop.

Now we address the parameter ranges for $\{y, M_{\Psi}, Q_{
\psi}\}$ in the parameter space.
First,
if one allows $\phi$ decaying into $\psi_{1,2}\bar{\psi}_{1,2}$, 
the total decay width for $\phi$ would be dominated by this channel,
which leads to a very small branching ratio $\text{Br}(\phi\rightarrow\gamma\gamma)$ typically of order $\leq 10^{-5}$ as a result of the fact,
\begin{eqnarray}\label{ratio}
\frac{\Gamma(\phi\rightarrow \gamma\gamma)}{\Gamma(\phi\rightarrow gg)} \sim \frac{1}{300} .
\end{eqnarray}
To account for the observed signal strength in Eq.(\ref{excess}),
we must forbid this decay channel, and impose $M_{\Psi}> m_{\phi}/2$.
Second, $\Gamma(\phi\rightarrow\gamma\gamma)/\Gamma(\phi\rightarrow gg)$ is roughly proportional to $Q_{\psi}^{4}$.
In order to obtain $\text{Br}(\phi\rightarrow\gamma\gamma)$ as large as possible,
one may choose large $Q_{\psi}$. 
Finally, a perturbative theory requires the Yukawa coupling constant $y\leq \sqrt{4\pi}$.
Based on the considerations above, 
we mainly focus on the following parameter ranges,
\begin{eqnarray}\label{ranges}
 375~\text{GeV} <M_{\Psi}< 2000~\text{GeV},~~~~Q_{\psi_{1}}=\{8/3, 5/3, 2/3, -1/3 \},~~~~ 0<y< \sqrt{4\pi}.
\end{eqnarray}

\subsection{Constraints}
Experimental constraints on $\phi$ mainly arise from limits at the 8 TeV LHC \cite{1506.02301,8tevcms, 8tevatlas,
8tev-diphoton, 8tev-zz, 8tev-ww, 8tev-dijet, 8tev-bb, 8tev-tt},
which are shown in Table \ref{csl}.
The $\gamma\gamma$, $Z\gamma$, $ZZ$, $WW$ and di-jet limits 
are shown in the red, green, purple, black and blue curve, respectively, 
above which the regions are excluded.

 \begin{table}[h!]
\begin{center}
\begin{tabular}{|c|c|c|c|c|}
  \hline
\text{cross sections}  &  \text{upper bounds (fb)} & \text{colors} \\
  \hline\hline
  $\sigma(pp\rightarrow \gamma\gamma) $ & $1.5$  &  \text{red} \\
\hline 
$\sigma(pp\rightarrow Z\gamma)$& $4$  & \text{green}  \\
\hline
$\sigma(pp\rightarrow ZZ)$&  $12$ & \text{purple} \\
\hline
$\sigma(pp\rightarrow WW)$&  $40$ &\text{black} \\
\hline
$\sigma(pp\rightarrow \text{di-jet})$&  $2500$ & \text{blue} \\
\hline
 \end{tabular}
\caption{Experimental limits on $\phi$ at 8 TeV LHC.}
\label{csl}
\end{center}
\end{table}

Experimental limits on vector-like quark $\psi$ are sensitive to its electric charge assignments.
If the electric charge $Q_{\psi_{1}}$ takes special values $\{8/3, 5/3,3/2,-1/3\}$,
in which case it allows mixing between $\psi_{1,2}$ and SM quarks,
decay channels such as $\psi_{1,2}\rightarrow \{tW, bW\}$ occur.
Otherwise, if $\psi_{1,2}$ takes some exotic electric value, 
which forbids the mixing effect, these limits can be obviously relaxed.
In this situation for $\psi_{1,2}$ pair produced at the LHC, 
they first hadronize into heavy ``mesons'', 
and then decay to SM final states.
In Table \ref{psi} we show the experimental limits at the 8 TeV LHC 
for different electric charges $Q_{\psi_{1}}$ and assumptions on its decay channel.

 \begin{table}[h!]
\begin{center}
\begin{tabular}{|c|c|c|c|c|}
  \hline
\text{charge}  &  \text{lower mass bound (GeV)} & assumption & \text{Refs} \\
  \hline\hline
$Q_{\psi_{1}}=8/3$ & $840$ &  $\text{Br}(\psi_{2}\rightarrow tW^{+})=100\%$  & \cite{1312.2391,1503.05425} \\
\hline
$Q_{\psi_{1}}=5/3$ & $920$ &  $\text{Br}(\psi_{2}\rightarrow tW^{+})=100\%$  & \cite{1509.04177} \\
\hline
$Q_{\psi_{1}}=2/3$ & $900$ &  $\text{Br}(\psi_{2}\rightarrow \{bZ, bH\})=100\%$  & \cite{1507.07129,1311.7667, 1209.1062} \\
\hline
$Q_{\psi_{1}}=-1/3$ & $800$ &  $\text{Br}(\psi_{1}\rightarrow \{bZ, bH\})=100\%$  & \cite{exotic4} \\
\hline
 \end{tabular}
\caption{Lower mass bounds on $M_{\Psi}$ at 8 TeV LHC for benchmark electric charge $Q_{\psi_{1}}=\{8/3,5/3,2/3,-1/3\}$. Note that in our case $\psi_{1}$ and $\psi_{2}$ have degenerate masses, and $Q_{\psi_{2}}=Q_{\psi_{1}}-1$.
The lower mass bound can be relaxed by adjusting the branching ratio.}
\label{psi}
\end{center}
\end{table}

\section{Results}
\begin{figure}
\centering
\begin{minipage}[b]{0.5\textwidth}
\centering
\includegraphics[width=3.2in]{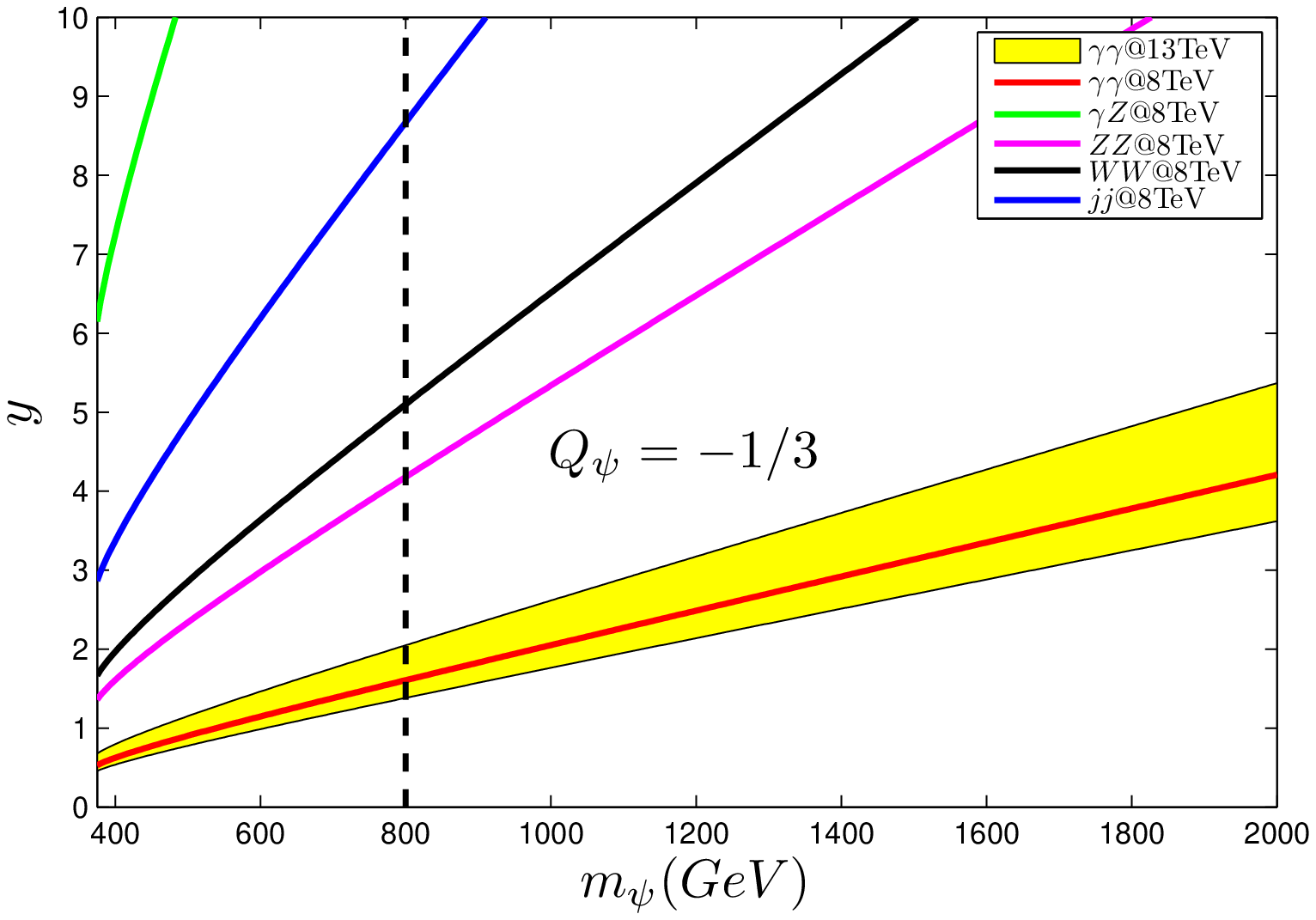}
\end{minipage}%
\centering
\begin{minipage}[b]{0.5\textwidth}
\centering
\includegraphics[width=3.2in]{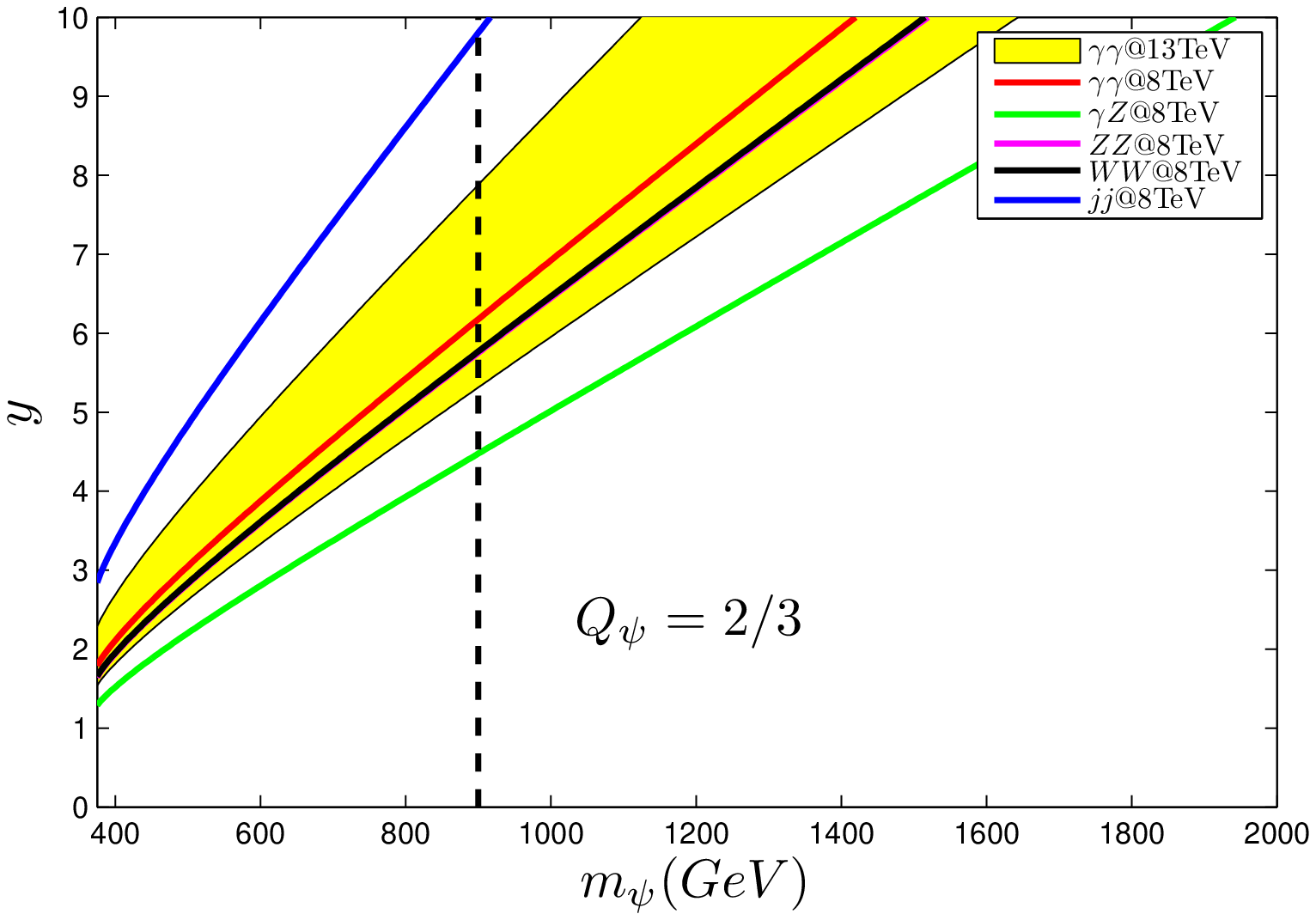}
\end{minipage}%
\\
\centering
\begin{minipage}[b]{0.5\textwidth}
\centering
\includegraphics[width=3.2in]{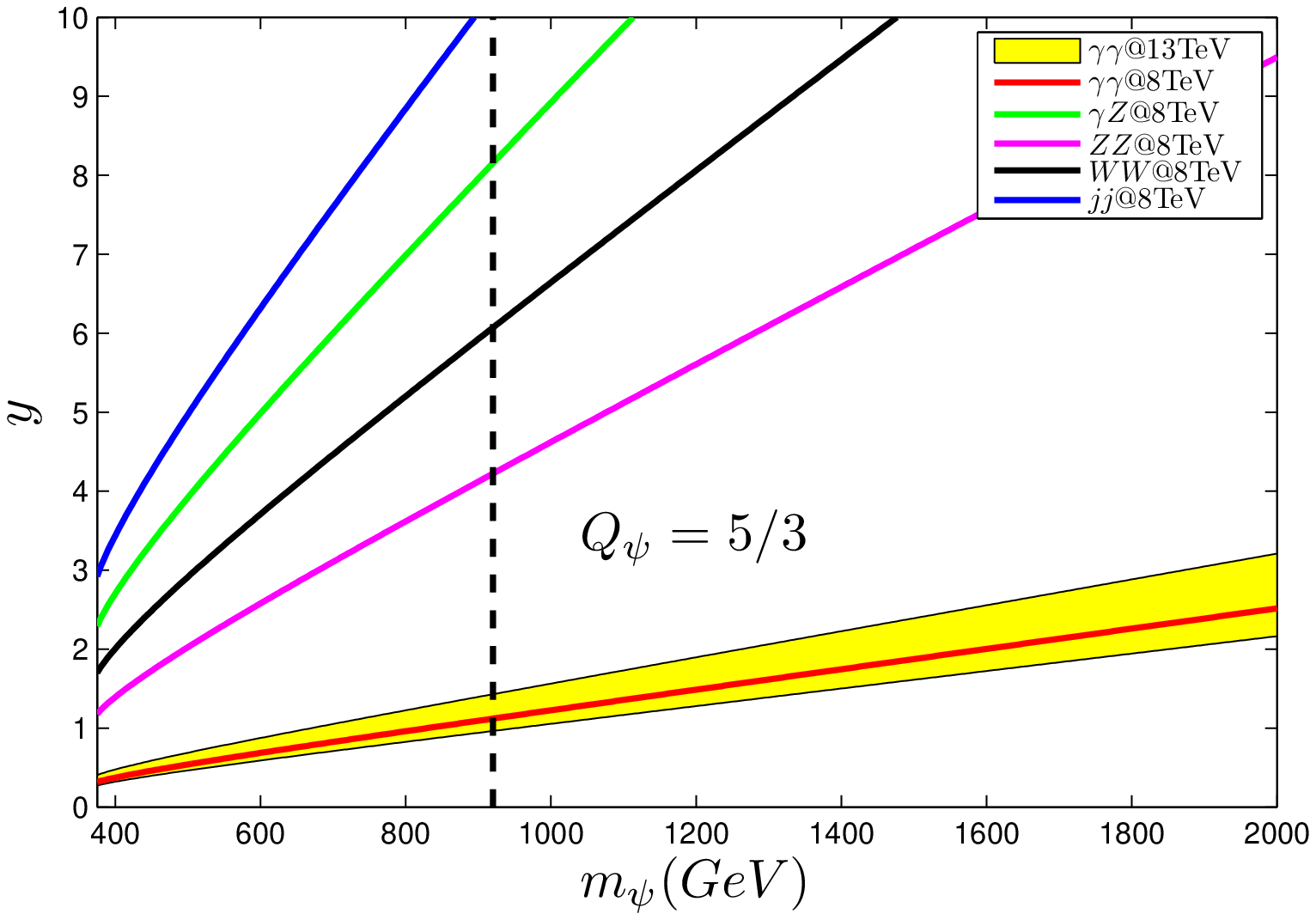}
\end{minipage}%
\centering
\begin{minipage}[b]{0.5\textwidth}
\centering
\includegraphics[width=3.2in]{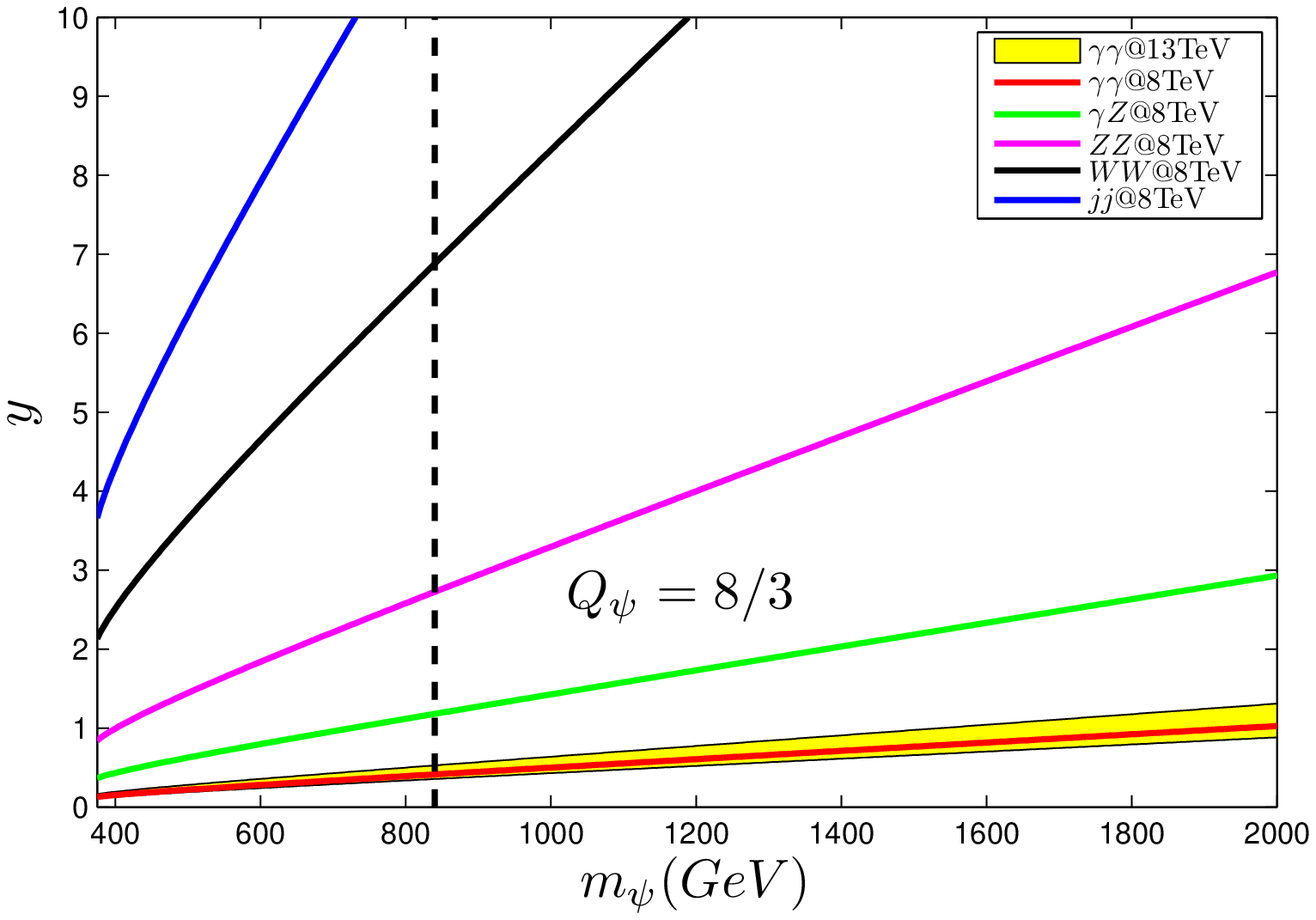}
\end{minipage}%
\caption{Yellow bands are the observed signal strength at 13 TeV LHC. 
Curves refer to experimental limits shown in Table \ref{csl}.
Each panel corresponds to a benchmark electric charge $Q_{\psi_{1}}=\{8/3, 5/3, 2/3,-1/3\}$, respectively. $M_{\Psi_{2}}$ is fixed to be 10 TeV. }
\label{scalar}
\end{figure}

\subsection{Scalar Resonance}
In this section we consider the SM singlet real scalar as the explanation of diphoton excess.
In Fig.\ref{scalar} the yellow bands correspond to the observed diphoton excess in the parameter space 
for fixed $M_{\Psi_{2}}=10$ TeV and four benchmark electric charges $Q_{\psi_{1}}=\{8/3, 5/3, 2/3,-1/3\}$.
In this figure, curves refer to experimental limits shown in Table \ref{csl},
above which regions are excluded.

$Q_{\psi_{1}}=\{-1/3, 5/3, 8/3\}$: 
For these three benchmark electric charges  
those regions in the yellow band below the the red solid curve
can explain the observed diphoton excess and 
are consistent with the experimental limits on $\phi$ in Table \ref{csl} simultaneously.
Furthermore, regions on the right hand of vertical dotted line survive 
after we take into account the constraints on vector-like quark in Table \ref{psi}.
Given the same $M_{\Psi}$ the Yukawa coupling constant $y$ as required to explain the diphoton excess tends to decrease as $\mid Q_{\psi_{1}}\mid$ increases.

$Q_{\psi_{1}}=2/3$: In contrast to the three benchmark electric charges above, 
 the model which actually corresponds to the 4th SM fermion generation is excluded by the $Z\gamma$ limit at  8 TeV LHC.
The reason arises from the fact that the production cross section for $\sigma(pp\rightarrow \phi)\cdot\text{Br}(\phi\rightarrow \gamma\gamma)$ 
is not a strictly monotonic function of $Q_{\psi_{1}}$.

\begin{figure}
\centering
\begin{minipage}[b]{0.5\textwidth}
\centering
\includegraphics[width=3.2in]{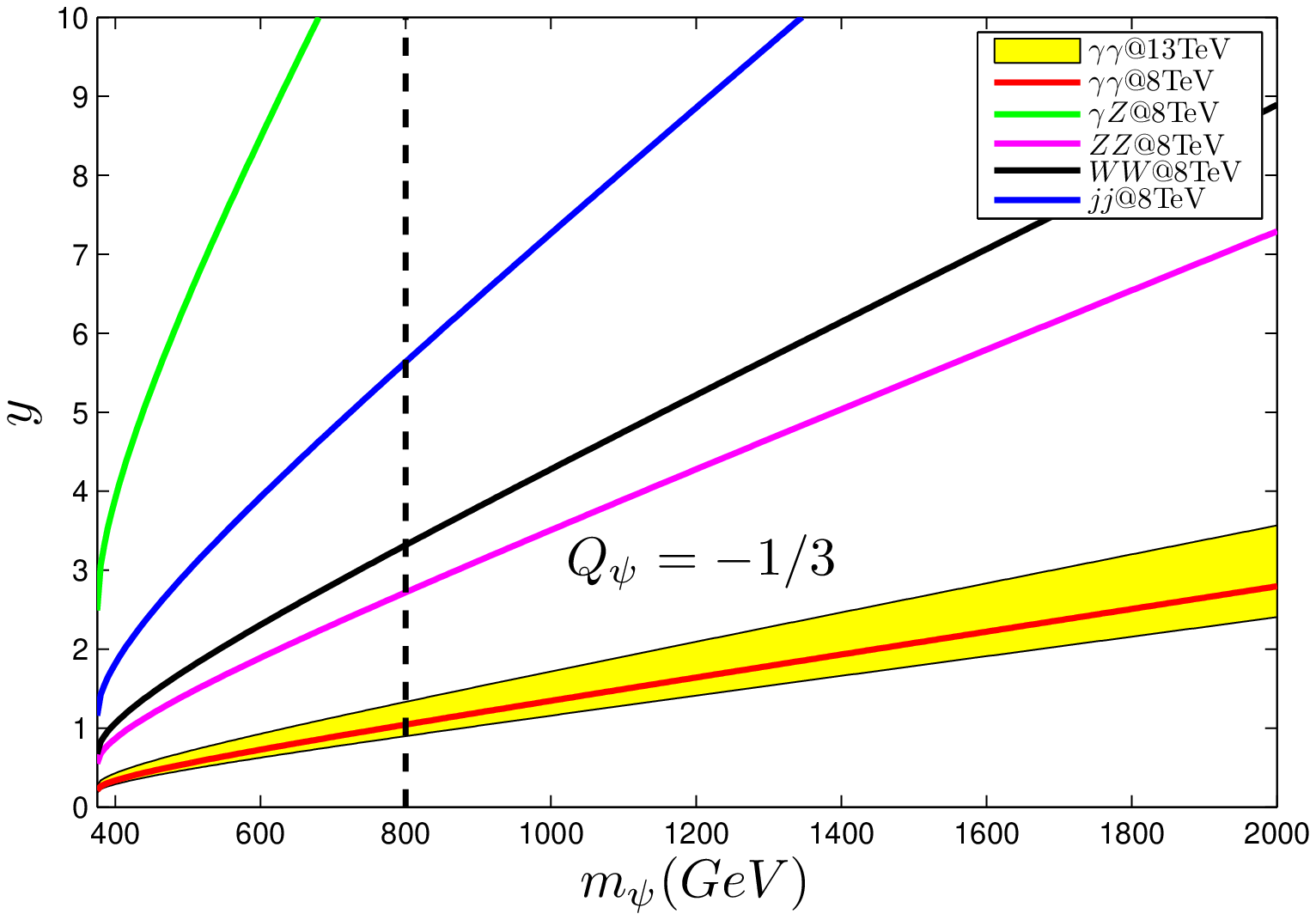}
\end{minipage}%
\centering
\begin{minipage}[b]{0.5\textwidth}
\centering
\includegraphics[width=3.2in]{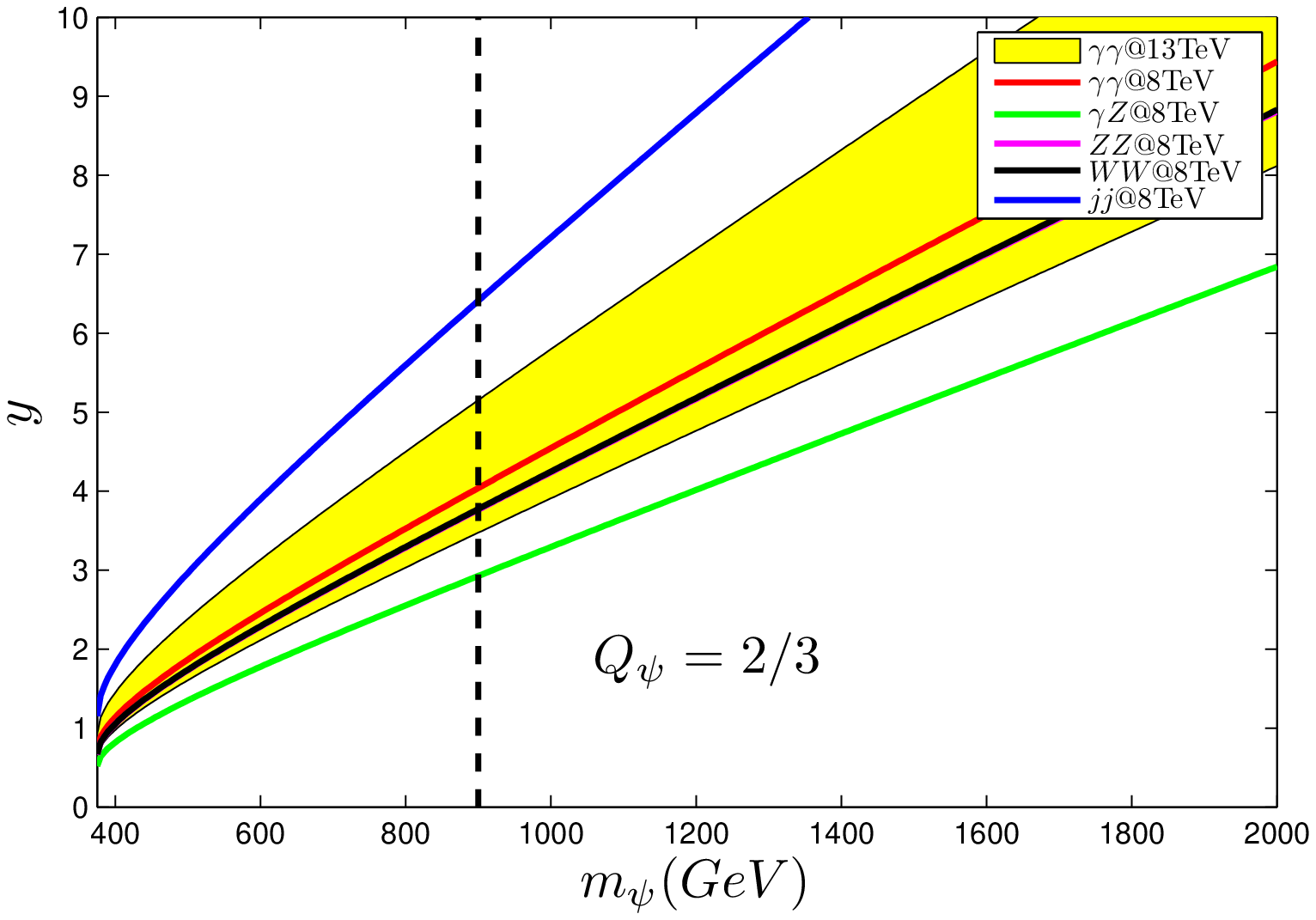}
\end{minipage}%
\\
\centering
\begin{minipage}[b]{0.5\textwidth}
\centering
\includegraphics[width=3.2in]{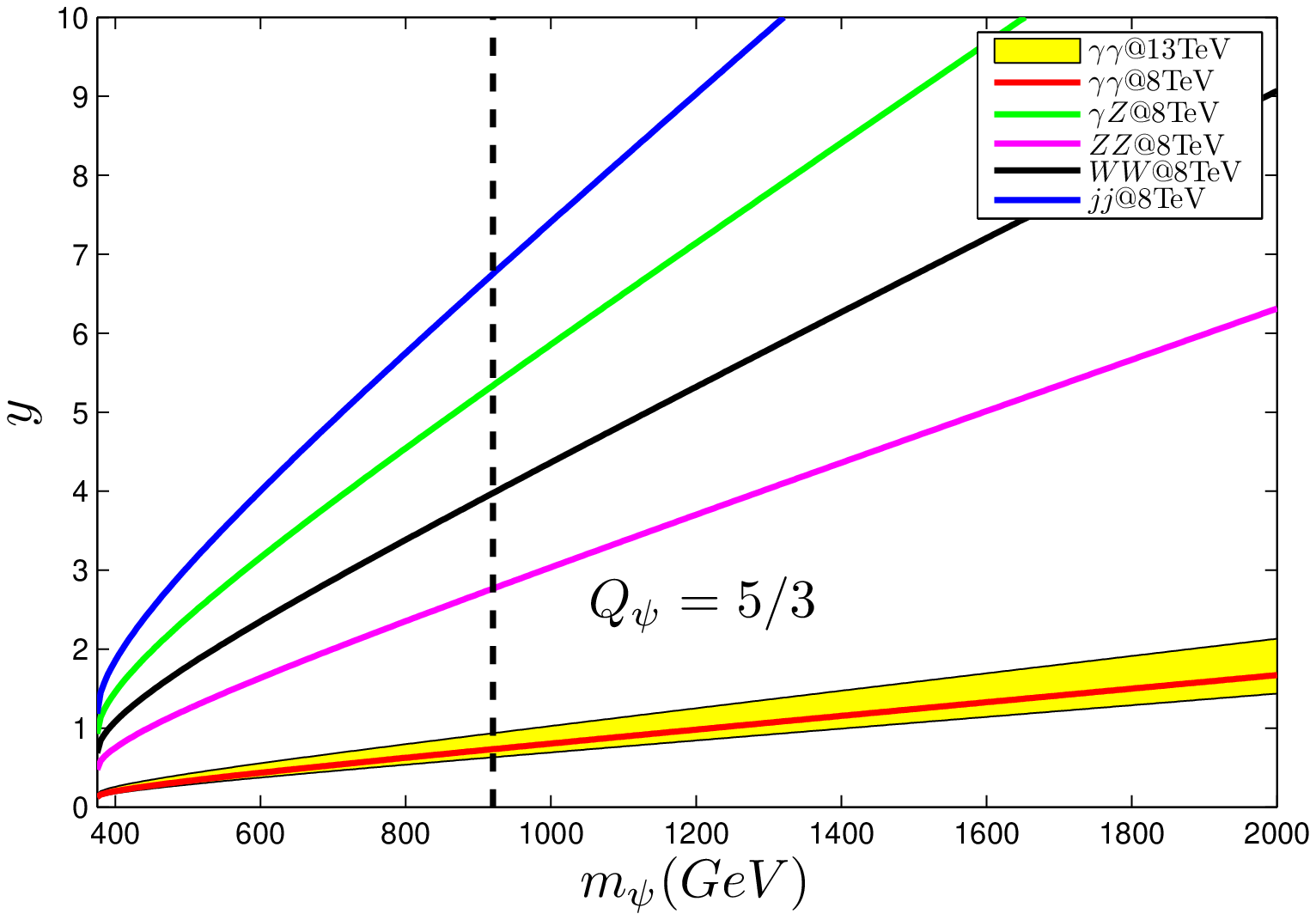}
\end{minipage}%
\centering
\begin{minipage}[b]{0.5\textwidth}
\centering
\includegraphics[width=3.2in]{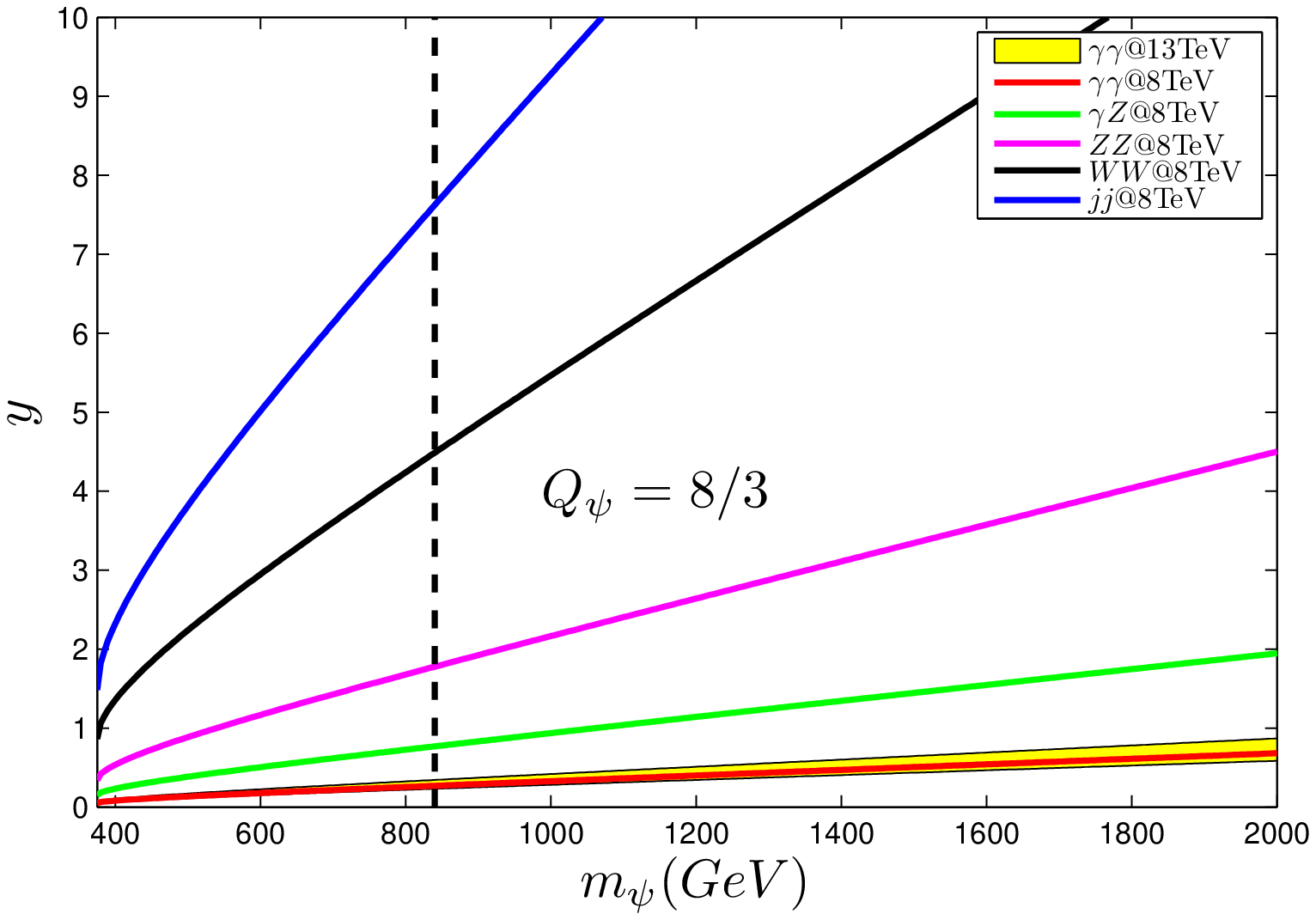}
\end{minipage}%
\caption{Similar to Fig.\ref{scalar} yellow bands correspond to the observed signal strength at 13 TeV LHC,
and curves refer to limits in Table \ref{csl}.}
\label{pscalar}
\end{figure}

\subsection{Pseudo-Scalar Resonance}\label{sec4}
Now we proceed to discuss the alternative explanation of diphoton excess via a SM singlet pseudo-scalar.
Similar to Fig.\ref{scalar},  the yellow bands in Fig.\ref{pscalar}  correspond to the diphoton excess.

$Q_{\psi_{1}}=\{-1/3, 5/3, 8/3\}$: Similar to the SM real scalar singlet,
there are viable parameter spaces for these three benchmark electric charges.
Fig.\ref{scalar} and Fig.\ref{pscalar} show that 
all of these parameter spaces are in the perturbative region,
which is an interesting feature for this model.
Our analysis indicates that the parameter spaces are not sensitive to the experimental limits in Table \ref{psi} in comparison with those in Table \ref{csl}. 
In Table \ref{wz13} we show the ranges for production cross sections 
$\sigma(pp\rightarrow \phi)\cdot\text{Br}(\phi\rightarrow \text{WW/ZZ})$ in the parameter spaces of Fig.\ref{scalar} and Fig. \ref{pscalar} at the 13 TeV LHC.

 \begin{table}
\begin{center}
\begin{tabular}{|c|c|c|c|}
  \hline
\text{channel}  &  $Q_{\psi_{1}}=-1/3$ & $Q_{\psi_{1}}=5/3$ & $Q_{\psi_{1}}=8/3$ \\
  \hline\hline
$\gamma\gamma$ & 5.0-6.8 &  5.0-6.8  & 5.0-6.8 \\
\hline
$\text{ZZ}$ & 6.1-8.2 &  2.9-3.9  & 1.0-1.3 \\
\hline
$\text{WW}$ & 13.5-18.3&  4.6-6.3  & 0.5-0.6 \\
\hline
 \end{tabular}
\caption{Production cross sections 
$\sigma(pp\rightarrow \phi)\cdot\text{Br}(\phi\rightarrow \text{WW/ZZ})$ in unit of fb in the parameter spaces of Fig.\ref{scalar} and Fig. \ref{pscalar} at the 13 TeV LHC. }
\label{wz13}
\end{center}
\end{table}

$Q_{\psi_{1}}=2/3$: Similar to the real scalar explanation in Fig.\ref{scalar},
vector-like quark with $Q_{\psi_{1}}=2/3$ is unable to explain the diphoton excess.

\section{Conclusion}
In this paper, we consider the possibilities that either a SM singlet scalar or singlet pseudo-scalar is responsible for the diphoton excess at 750 GeV in the 13 TeV LHC.
To analyze the allowed parameter space, 
we take the experimental limits at 8 TeV LHC.
For the four benchmark electric charges $Q_{\psi_{1}}=\{-1/3, 2/3, 5/3, 8/3\}$
we find that there are viable parameter spaces except for $Q_{\psi_{1}}=2/3$.
Moreover, all of these viable parameter spaces are in the perturbative region,
which differ from some attempts in Ref.\cite{1512.04850}- Ref.\cite{1512.06113} to address the diphoton excess.
Given the fact that branching ratios for decays $\phi\rightarrow \{\gamma\gamma,ZZ, WW\}$ are all of the same order,
signal excesses in $pp\rightarrow\phi\rightarrow WW$ and $pp\rightarrow\phi\rightarrow ZZ$ will be exposed 
for larger integrated luminosity.\\

~~~~~~~~~~~~~~~~~~~~~~~~~~~~~~~~~~~~
$\mathbf{Acknowledgments}$\\
We thank A. Strumia for correspondence and the referee for useful suggestions.
This work is supported in part by the National Natural Science Foundation of China under grant No.11205255 and 11405015.

\end{document}